\documentclass[referee]{aa}

\usepackage{graphicx}

\usepackage{txfonts}
\begin{document}
   \title{Investigation of Dynamics of Self-Similarly Evolving Magnetic Clouds}

   \author{Giorgi Dalakishvili
          \inst{1,4}
          \and
          Andria Rogava\inst{2,3}
          \and
    Giovanni Lapenta\inst{4}
     \and
    Stefaan Poedts\inst{4}
}

   \institute{
Institut f\"ur Weltraum und Astrophysik, Ruhr-Universit\"at Bochum (RUB)\\
              \email{giorgi@tp4.rub.de}
         \and
Centre for Theoretical Astrophysics, ITP, Ilia State University\\
             \email{andria.rogava@iliauni.edu.ge}
     \and
Abdus Salam International Centre for Theoretical Physics
     \and
Centre for Plasma Astrophyics, Katholieke Universiteit Leuven \\
       \email{Giovanni.Lapenta@wis.kuleuven.be},
        \email{Stefaan.Poedts@wis.kuleuven.be}
             }

   \date{Received,Accepted }

  \abstract
   {Magnetic clouds (MCs) are ``magnetized plasma clouds" moving in the
   solar wind. MCs transport magnetic flux and helicity away from the Sun. These structures are not
   stationary but feature temporal evolution. Commonly, simplified MC models are considered.}
   {The goal of the present study is to investigate the dynamics of more general, \textit{radially expanding} MCs.
    They are considered as cylindrically symmetric magnetic structures with low plasma $\beta$.}
   {The self-similar approach method and a numerical approach are used.}
   {It is shown that the forces are balanced in the considered self-similarly evolving, cylindrically symmetric
    magnetic structures. Explicit analytical expressions for magnetic field, plasma velocity, density
    and pressure within MCs are derived. These solutions are characterized by conserved values of magnetic
    flux and  helicity. We also investigate the dynamics of self-similarly evolving MCs by means of
    the numerical code ``Graale''. In addition, their expansion in a medium with higher density and higher
    plasma $\beta$ is studied. It is shown that the physical parameters of the
    MCs maintain their self-similar character throughout their evolution.}
   {A comparison of the different self-similar and numerical solutions allows us to conclude that the evolving MCs are quite
    adequately described by our self-similar solutions - they retain their self-similar, coherent nature for quite a long time
    and over large distances from the Sun.}


\titlerunning{Investigation of Dynamics of Self-Similarly Evolving Magnetic Clouds}

\authorrunning{Dalakishvili et al.}
   \maketitle

\section{Introduction}

It is well-known that \textit{coronal mass ejections} (CMEs) are one
of the most significant forms of solar activity. They carry enormous
masses of plasma threaded by the magnetic field away into the
interplanetary medium. Further away from the Sun, these large-scale,
dynamical plasma structures are commonly called
\textit{interplanetary coronal mass ejections} (ICMEs). Magnetic
clouds (MCs) form a subset of ICMEs (Klein \& Burlaga \cite{kle82},
Burlaga \cite{bur91}, Farrugia et al.\ \cite{far95}). Spacecrafts
crossing the central parts of such MCs provide valuable information
about their physical characteristics. It turns out that MCs have a
strong magnetic field, low proton temperatures (low plasma $\beta$,
compared to the ambient solar wind with the same speed) and they
feature a substantial and smooth rotation of the magnetic field
vector. These three features of MCs are selected as signatures of
MCs (Nakwacki et al.\ \cite{nak08}). The MCs are also characterized
by a coherence of the magnetic field (low level of fluctuations).
The radial dimension of \textbf{a MC} is typically $\approx
0.25\;$AU (at $1\;$AU).

These \textit{in situ} observations of the physical properties of
MCs are considered as important steps towards the prediction of the
geophysical effectiveness of their interaction with the Earth's
magnetosphere, for space weather forecasts and related issues.

Different models for the structures of magnetic clouds have been
proposed. There is no general agreement about the large scale
structure of MCs. Commonly, the local structure of MCs is considered
in the form of cylindrically symmetric force-free configurations
(Burlaga \cite{bur88}, \cite{bur91}, Demoulin \& Dasso
\cite{dem09}). It is often suggested that the ends of MCs connect to
the surface of the Sun while, according to other models, MCs are
described as tori (Vandas et al. \cite{van06,van09}, Romashets et
al. \cite{RVP06,RVP07}). In a number of studies, MCs are considered
as force-free, static, axially symmetric flux ropes and their
magnetic field is constructed on the basis of Lundquist's model
(Burlaga \cite{bur88}, Lepping et al. \cite{lep90}, Farrugia et al.
\cite{far93}). Observations show, however, that MCs do not stay
static but expand while propagating in the solar wind and they keep
expanding well beyond 1~AU (Burlaga \cite{bur91}, Demoulin
\cite{dem08}, Demoulin \& Dasso \cite{dem09}, Bothmer \& Schwenn
\cite{bot98}). In a large majority of the cases it is observed that
the frontal parts of the MCs propagate with higher velocities than
their back regions. This shows that, with respect to the MC's own
cylindrical set of coordinates, the radial size of those cylindrical
MCs increases (Nakwacki et al.\ \cite{nak08}). Theoretical models
including the effect of radial expansion have been proposed before
(Osherovich et al. \cite{osh95}, Farrugia et al. \cite{far93},
Nakwacki et al. \cite{nak08}). In these models, only the radial
expansion is taken into account and solutions have been found for
all plasma parameters. There  are other studies (Shimazu \& Vandas
\cite{shi02}, Demoulin \& Dasso \cite{dem09}), however, where the
axial expansion is also included.

Previous studies also showed that inside MCs the density
\textbf{drops as $d^{-2.4}$ (Bothmer \& Schwenn \cite{bot98}), i.e.
the} volume of MC increases as $d^{2.4}$, where $d$ denotes the
distance from the Sun. The radius of the MCs, denoted by $(R)$, also
increased and at a rate changing with the distance, viz.\  as: $R
\sim d^{0.8}$ (Bothmer \& Schwenn \cite{bot98}). Since the surface
of the MC's cross section perpendicular to its axis increases as
$R^{2}\sim d^{1.6}$, and the MC's volume increases as $d^{2.4}$, the
MC's longitudinal size should increase as $d^{0.8}$. Therefore,
according to Bothmer \& Schwenn's data, the MCs are radially
expanding and also show an extension along their axis.

In the present study, we consider self-similarly expanding
cylindrical MCs that are able to expand both in the radial  and
longitudinal directions. We consider the problem in the frame of the
MC and in cylindrical coordinates related with the MC, i.e.\ with a
longitudinal axis $Z$ that coincides with the MC's axis. Overall
cylindrical symmetry of the MC is assumed. Based on these
assumptions, we derive the appropriate full set of non-stationary
MHD equations and find their analytical solutions. The logical and
natural consequence of the assumptions of self-similarity and
cylindrical symmetry is that the dynamic forces acting upon the MCs
are balanced. The solutions include expressions for the plasma
magnetic field, velocity, mass density and thermal pressure.

An important feature of our model is that certain significant
characteristics of the MCs
--- magnetic flux and helicity
--- are conserved. We separately consider also the particular case
of a MC that is allowed to expand only in the radial direction. It
can be shown that in this case, the MHD equations do not have any
physical, self-similar solution.

\section{Self-similar expanding MC models}
\subsection{General equations and self similar expansion}
 In order to perform an analytic study of the dynamics of
magnetic clouds, we have to start from the full set of MHD
equations:
$$
\bigtriangledown\cdot\textbf{B}=0, \eqno(1)
$$
$$
{\partial}_t{\textbf{B}}= \nabla\times\left[\textbf{V}\times\textbf{B}\right], \eqno(2)
$$
$$
{\partial}_t{\varrho}+\nabla\cdot\left(\varrho\textbf{V} \right)=0,
\eqno(3)
$$
$$
\varrho[{\partial}_t+(\textbf{V}\cdot\nabla)]\textbf{V}=(1/4
\pi)(\bigtriangledown\times\textbf{B})\times\textbf{B}-\nabla\cdot
p, \eqno(4)
$$

In these equations, $p$ denotes the thermal plasma pressure,
$\varrho$ is the density, ${\bf V}$ is the velocity field and
$\textbf{B}$ denotes the magnetic field.

In a number of previous studies, the MCs were considered as
cylindrical magnetic structures, characterized by axial symmetry. In
the present consideration, both symmetry along the $Z$ axis
(${\partial}_z=0$) and the azimuthal symmetry
(${\partial}_{\varphi}=0$) are assumed. The axially-symmetric
magnetic field can then be expressed in the following way
$$
\textbf{B} \equiv [0,~~ B_{\varphi}, ~~B_{z}], \eqno(5)
$$
where $B_{\varphi}=B_{\varphi}(r,t)$ and $B_{z}=B_{z}(r,t)$. Note
that this representation satisfies the solenoidal condition.

The self-similar approach, adopted here, implies that the temporal
evolution of the physical functions is controlled by the following
self-similar variable:
$$
\xi=\frac{r}{\Phi(t)},\eqno(6)
$$
where $\Phi(t)$ denotes a function of time. Let us search solutions
of the MHD equations in the following form
 (in analogy with Low \cite{low82}):

$$B_{\varphi}=\Phi^{\delta}Q_{\varphi}(\xi),\eqno(7a)$$

$$B_{z}=\Phi^{\sigma}Q_{z}(\xi),\eqno(7b)$$

$$
\varrho=\Phi^{\alpha}\tilde \rho(\xi),\eqno(7c)
$$
$$
p=\Phi^{\beta}\tilde p(\xi),\eqno(7d)
$$

One can see that, the type of solutions introduced by Eqs.~(7a-7d)
evolve self-similarly and are characterized by a particular time
scaling.

Here  $Q_{\varphi},$ $Q_{z},$ $\tilde \rho$ and $\tilde p$ are
functions of the self similar variable $\xi$. $\Phi^{\delta}$,
$\Phi^{\sigma}$, $\Phi^{\alpha}$ and $\Phi^{\beta}$ show the time
scaling of the azimuthal and  longitudinal components of the magnetic
field, the plasma density and the plasma pressure, respectively.

\subsection{Solution of the induction equation}

We consider both a radial and a longitudinal expansion of the MC but no motion in
the azimuthal direction is considered. In this case the
\textit{Eulerian} velocity field of the plasma, $\textbf{V}$,  can be
expressed in the following way:
$$
\textbf{V} = [V_{r},~~0,~~V_{z}]. \eqno(8)
$$

Here, we assume that the radial component of the velocity
$V_{r}=V_{r}(r,t)$, and the $z-$component $V_{z}=V_{z}(z,t)$, i.e.\
we assume that the MC maintains its cylindrical shape during its
evolution.

After substitution of  Eq.~$(8)$ in Eq.~$(2)$ we derive:
$$
\partial_t B_{z} + \frac{1}{r}\partial_r (r V_{r} B_{z}) = 0, \eqno(9a)
$$
After taking into account relations (6), (7b) and the relations (A1)
and (A2) given in the appendix, Eq.~(9a) can be rewritten as
follows:

$$
Q_{z}\left[\sigma\dot\Phi+\frac{V_{r}}{\xi}+\Phi\partial_r
V_{r}\right]+Q_{z}'\left[V_{r}-\xi\dot\Phi\right]=0. \eqno(9b)
$$

Here $Q_{z}'$ corresponds to $dQ_{z}(\xi)/d\xi$. Equation (9a)
(therefore equation (9b)) is satisfied for arbitrary $Q_{z}$ only
when:
$$
V_{r}-\xi\dot\Phi=0, \eqno(10a)
$$
and

$$
\sigma\dot\Phi+\frac{V_{r}}{\xi}+\Phi\partial_r V_{r}=0. \eqno(10b)
$$

From Eq.~(10a) and Eq.~(10b) follows that radial component of the
Eulerian plasma velocity is described as follows:
$$
V_{r}=r \dot{\Phi}/{\Phi}, \eqno(11)
$$

and
$$
\sigma=-2, \eqno(12)
$$
 where $\Phi$ is the function of time mentioned in Eq.~(6).

One can check that for $\sigma=-2$ the longitudinal magnetic flux
$\phi_{z}$ is conserved. Nakwacki et al. (\cite{nak08}) analyzed
different MC models and derived expressions for the magnetic flux,
the magnetic helicity and the magnetic energy per unit length along
the flux tube. The models which are in good agreement with
observations are characterized by the conservation of $\phi_{z}$,
see also Berdichevsky et al. (\cite{ber03}).

Let us analyze the $\varphi$-component of the induction equation,
Eq.~(2):

$$
\partial_t({B_{\varphi}}) + {B}_{\varphi} \partial_z V_{z} + \partial_r(V_{r} {B_{\varphi}}) = 0. \eqno(13)
$$

The combination of Eqs.~(11) and (13) leads to the following
important relation:
$$
(\delta+1)\dot{\Phi}/{\Phi}+\partial_z V_{z} = 0. \eqno(14)
$$

After taking into account expressions (5) and (8) in combination
with the assumption of azimuthal symmetry, one can see that radial
component of the induction equation, Eq.~(2), is automatically
satisfied and does not lead to any additional restrictions.

\subsection{Self-similar solutions}

After inserting the expression for the plasma density (7c), together
with the velocity from Eq.~(8), with Eq.~(11) for the radial
component, in the mass conservation law Eq.~(3), we obtain another
important relation, viz.\
$$
(\alpha+2)\dot{\Phi}/{\Phi}+\partial_z V_{z} = 0. \eqno(15)
$$

Obviously, in order to have consistency between Eqs.~(14) and (15),
one should have: $\alpha+2=\delta+1$.

The $z$-component of the equation of motion, Eq.~(4), helps to derive
an expression for the $z$-component of the plasma velocity:
$$
\partial_t V_{z}+V_{z}\partial_z V_{z}=0. \eqno(16a)
$$

Let us try to solve the partial differential equation (16a) by using
the variable separation technique, i.e.\ we assume that

$$V_{z}(z,t)=Z(z)T(t).\eqno(16b)$$

Substitution of expression (16b) in Eq.~(16a) yields:
$$
\dot TZ+ZT^{2}Z'=0,\eqno(17a)
$$

here $\dot{F}\equiv\partial_tF$ denotes the first order time
derivative of a function $F$. Hereafter we will use, for indicating
second order derivatives, the notation: $\ddot{F} \equiv
\partial_t^2F$. While $Z'$ stands for $dZ/dz$.

It follows from Eq.~(17a) that:
$$
-\frac{\dot T}{T^{2}}=Z'=const,\eqno(17b)
$$
Equation~(17b) can be decomposed into two ODEs, viz.\
$$
-\frac{\dot T}{T^{2}}=\lambda,\eqno(17c)
$$
and
$$
Z'=\lambda.\eqno(17d)
$$
Here, $\lambda$ is an arbitrary constant.

After solving the ODEs (17c-17d)  with the assumption that, at the
surface $z=0$,  $V_{z}=0$, we derive the following expressions:
$$
T=\frac{T_{0}}{1+\lambda T_{0}t},\eqno(18a)
$$
and:
$$
Z=\lambda z,\eqno(18b)
$$
where $\lambda$ and $T_{0}$ are constants.

After inserting Eqs.~(18a-18b) in Eq.~(16b), we obtain the wanted
expression for $V_{z}$:
$$
V_{z}=\frac{zk}{1+kt}.\eqno(19)
$$
Here, $k\equiv\lambda T_{0}.$

We assumed that locally the MC could be described as a cylindrical
structure. Let us investigate the evolution of the length $L$ of
this cylindrical structure. For this purpose, let us describe the
temporal evolution of the $z$-coordinate of the plasma element
located at the position $z=L$ at time $t$. The Lagrangian velocity
of this element coincides with the Eulerian velocity of the plasma
flow at time $t$ \textbf{and $z=L$.  If} at a certain time the
coordinate of this element is $L,$ then its Lagrangian velocity is:

$$
V_{L}=\frac{d L}{dt}.\eqno(20a)
$$
From Eq.~$(19)$ we then have:
$$
\frac{d L}{dt}=\frac{Lk}{1+kt}.\eqno(20b)
$$
The solution of this ordinary differential equation (20b) gives the
following expression for the \textbf{longitudinal size of the
considered cylindrical structure}:

$$
L=L_{0}(1+kt),\eqno(20c)
$$
\textbf{where $L_{0}$ is the length of the cylinder at $t=0$.}

The radial component of the equation of
motion, in combination with the expressions for the magnetic field,
the velocity and the plasma density leads to:
$$
\Phi^{\alpha}\xi\ddot{\Phi}\tilde\rho=F_{r}. \eqno(21)
$$
Here, $F_{r}$ denotes the radial component of the total force. In
terms of $\tilde p$, $Q_{\varphi,z}$, and $\xi$, this force
component can be expressed as (details of the derivation are given
in the appendix):
$$
F_{r}=
-\frac{1}{4\pi}\left[\frac{1}{\Phi^{5}}Q_{z}'Q_{z}+\Phi^{\delta-3}\left(\frac{Q_{\varphi}^{2}}{\xi}+Q_{\varphi}'Q_{\varphi}\right)\right]-\Phi^{\beta-1}\tilde
p',\eqno(22)
$$
where
 $Q_{\varphi,z}^{/} = dQ_{\varphi,z}/d\xi$, and $\tilde
p'=d\tilde p/d\xi$. In order to have a self-consistent time scaling
for all terms in Eq.~(22), one has to require that $\delta=-2$ and
$\beta=-4$. From a comparison of Eq.~(14) to Eq.~(15), follows that
if $\delta=-2$, then $\alpha=-3$.

At the same time, from Eq.~(15) and Eq.~(19):
$$
\frac{\dot\Phi}{\Phi} = \frac{k}{1+kt}. \eqno(23)
$$
Equation~(23) is an ordinary differential equation in terms of
$\Phi(t)$. After solving this ODE, we find the following explicit
expression for $\Phi(t)$:
$$
\Phi = \Phi_{0}(1+kt).\eqno(24)
$$
Here $\Phi_{0}$ is a constant parameter.

 The substitution of expression
(24) in Eq.~(21) leads to an important conclusion: calculating the
magnetic and pressure gradient forces, we see that for the
self-similarly evolving, cylindrical, axially-symmetric structure
the magnetic force, $F_m \equiv 1/(4
\pi)(\bigtriangledown\times\textbf{B})\times\textbf{B}$ and the
thermal pressure gradient force, $F_p \equiv -\bigtriangledown\cdot
p$ are exactly balanced, i.e.\
$$
F_{r}=F_m + F_p =0.\eqno(25)
$$

If we associate the value $\xi_{0}$ of the self-similar variable
$\xi$ with the boundary of the MC, then the expression of the MC
Lagrangian velocity is given by (Low \cite{low82}):
$$
V_{s}=\frac{dR}{dt} = \xi_{0}\frac{d\Phi}{dt}. \eqno(26)
$$
After substitution of the expression (24) for $\Phi$ in Eq.~(26), we
can derive a time-dependent solution for the MC radius:
$$
R=R_{0}(1+kt). \eqno(27)
$$
Note that the form of this expression coincides with the one given
by Nakwacki et al. (\cite{nak08}).

\subsection{Plasma and force-free field evolution}

The rest of the solutions readily follows from the derived equations,
yielding:
$$
V_{r}=\frac{rk}{1+kt},\eqno(28a)
$$
$$
\varrho=\frac{\tilde \varrho}{(1+kt)^{3}},\eqno(28b)
$$
and
$$
p=\frac{\tilde p}{(1+kt)^4}.\eqno(28c)
$$
Here, $\tilde\varrho$ and $\tilde p$ are arbitrary functions of
$\xi=r/\Phi$.

After analysis of the expressions for pressure and density
(28b-28c), one can check that, for systems characterized by entropy
conservation, the entropy conservation law is satisfied only if the
polytropic index $\gamma=4/3$. Actually, this is a common feature of
all different self-similar systems (Low \cite{low82}, Farrugia et
al. \cite{far95}, Finn et al. \cite{fin04}).

From various observations it is known that MCs are characterized
with low plasma $\beta$'s (Burlaga et al. \cite{bur81}, Burlaga
\cite{bur91}, Bothmer \& Schwenn \cite{bot98}). The thermal pressure
term in the total force could be neglected and this implies that we
have to construct a force-free magnetic field that evolves in a
self-similar way. The cylindrically symmetric force-free structure
of the MC's magnetic field is indeed advocated by a number of
researches (Burlaga \cite{bur88}, Lepping et al. \cite{lep90},
Farrugia et al. \cite{far93}, Farrugia et al. \cite{far95}, Nakwacki
et al. \cite{nak08}, Demoulin \& Dasso \cite{dem09}). A force-free
magnetic field satisfies the following relation:
$$
\bigtriangledown\times\textbf{B}=\mu\textbf{B}.\eqno(29)
$$
If we rewrite the vectorial equation (29) for each component of
vectors, taking in to account expressions (7a), (7b) and (12), we
obtain:
$$
-Q_{z}'=\mu\Phi Q_{\varphi},\eqno(30a)
$$
$$
Q_{\varphi}'+\frac{Q_{\varphi}}{\xi}=\mu\Phi Q_{z},\eqno(30b)
$$
here $Q_{\varphi,z}'$ stands for $dQ_{\varphi,z}/d\xi$. If we take
the derivative of both terms of Eq.~(30a) with respect to the
variable $\xi$, we get:
$$
-Q_{z}''=\mu\Phi Q_{\varphi}'.\eqno(30c)
$$
\textbf{Here it was assumed that $\mu$ does not depend on $\xi$. In
general, however, $\mu$ could be a function of $\xi$.}

If we take in to account expressions (30a) and (30c),  we can derive
from Eq.~(30b) an ordinary differential equation for $Q_{z}$:
$$
Q_{z}''+\frac{Q_{z}'}{\xi}+\mu^{2}\Phi^{2}Q_{z}=0.\eqno(31a)
$$
With the following transformation of variables: $x=\mu\Phi\xi$, we
can rewrite Eq.~(31a) as follows:
$$
\frac{d^{2}Q_{z}}{dx^{2}}+\frac{1}{x}\frac{dQ_{z}}{dx}+Q_{z}=0.\eqno(31b)
$$

Actually Eq.~(31b) is a Bessel equation of zero order, with the
following solution:
$$
Q_{z}=J_{0}(x)=C_{0}J_{0}(\mu\Phi\xi),\eqno(32a)
$$
where $J_{0}(x)$   is the Bessel function of the first kind, $C_{0}$
is a constant parameter. Notice that the solution which is not
characterized with a singularity at $x=0$ has been chosen. The
substitution of Eq.~(32a) in relation (30a) leads to an expression
for $Q_{\varphi}$:
 $$
 Q_{\varphi}=J_{1}(x)=C_{0}J_{1}(\mu\Phi\xi),\eqno(32b)
 $$
with $J_{1}(x)$ the Bessel function of the first kind.

\textbf{From Eqs.~(32a-b) we see that $Q_{\varphi}$ and $Q_{z}$ are
the functions of $\mu\Phi\xi$}.  \textbf{Since we assumed above
that} $\mu$ \textbf{is not a function of $\xi$, that $\Phi$ is only
a function of $t$ (Eq.~(6)), and  that  $Q_{\varphi,z}$ are
functions of only $\xi$ (Eqs.~(7a,7b)), it follows that
$\mu\Phi=const.$} \textbf{The substitution of Eqs.~(32a), (32b),
(24) and (6) in Eqs.~(7a) and (7b) respectively, taking into account
that $\delta=\sigma=-2$, leads to the following expressions for
 the components of the magnetic field:}
$$
B_{r}=0,\eqno(33a)
$$
$$
B_{\varphi}=\frac{B_{0}}{(1+kt)^{2}}J_{1}\left(\frac{r}{r_{0}(1+kt)}
\right)\textbf{,}\eqno(33b)
$$
and
$$
B_{z}=\frac{B_{0}}{(1+kt)^{2}}J_{0}\left(\frac{r}{r_{0}(1+kt)}
\right),\eqno(33c)
$$

\textbf{where  $B_{0}$ and $r_{0}$ are constants.}

\textbf{($C_{0}/\Phi_{0}^{2}$ is changed by $B_{0}$ and
$\mu\xi/\Phi_{0}$ is substituted by $1/r_{0}$.)}

From  Eqs.~$(33a-c)$ we can calculate important expressions for the
magnetic flux and the helicity (Nakwaci et al. \cite{nak08})
associated with the MC:
$$
\Phi_{z}=\frac{2\pi}{\chi}R\frac{B_{0}}{(1+kt)^{2}}J_{1}(\chi
R),\eqno(34a)
$$
$$
\Phi_{\varphi}=\frac{1}{\chi}\frac{B_{0}}{(1+kt)^{2}}L(1-J_{0}(\chi
R)),\eqno(34b)
$$
and
$$
H=\frac{2\pi}{\chi}R^{2}\frac{B_{0}^{2}}{(1+kt)^{4}}L(J_{1}^2(\chi
R)-J_{0}(\chi R)J_{2}(\chi R)+J_{0}^2(\chi R)),\eqno(34c)
$$
where $\chi \equiv 1/(r_{0}(1+kt))$.

By its physical meaning $\Phi_{z}$ is the magnetic flux across the
surface perpendicular to the axis of a MC, while $\Phi_{\varphi}$ is
the magnetic flux across the surface defined by the magnetic axis
and the radial direction. Moreover, $R$ denotes the radius of the MC
and $L$ is longitudinal length of the cylindrical structure. By
inserting in Eqs.~$(34a-c)$ the corresponding expressions for $R$
and $L$ we find:
$$
\Phi_{z}=2\pi R_{0}r_{0}B_{0}J_{1}\left(
\frac{R_{0}}{r_{0}}\right)=const,\eqno(35a)
$$
$$
\Phi_{\varphi}=B_{0}r_{0}L_{0}\left[ 1-J_{0}\left(
\frac{R_{0}}{r_{0}}\right)\right]=const,  \eqno(35b)
$$
and also
$$
H=2\pi r_{0}R_{0}^{2}B_{0}^{2}L_{0}\left[ J_{1}^2\left(
\frac{R_{0}}{r_{0}}\right) - J_{0}\left( \frac{R_{0}}{r_{0}}\right)
J_{2}\left( \frac{R_{0}}{r_{0}}\right) +J_{0}^2\left(
\frac{R_{0}}{r_{0}}\right)\right]=const. \eqno(35c)
$$

From  these results \textbf{it follows that the obtained solutions}
ensure the conservation of magnetic flux and helicity inside the
cylindrical MC described by our model.

\section{Radially expanding MCs}

The purpose of this section is to find solutions for the physical
variables in the case where only the radial size of the MC
increases. One can see that the solutions in this case do not remain
self-similar, although initially a self-similar expansion is assumed
in the radial direction.

\textbf{Below we  consider MCs} that are expanding only radially;
i.e.\ with $V_{z}=0$. In this case Eq.~(14) implies $\dot\Phi=0$ or
$\delta=-1$. The case with $\dot\Phi=0$ corresponds to the
\textit{stationary} state, which is trivial. Let us consider the
case when $\dot\Phi\neq 0$ but $\delta=-1$. In order to provide a
consistent time-scaling of all terms in Eq.~(22), we have to
satisfy:
$$
Q_{\varphi}Q_{\varphi}'+\frac{Q_{\varphi}^2}{\xi}=\frac{Q_{\varphi}}{\xi}\partial_\xi
(\xi Q_{\varphi})=0. \eqno(36a)
$$

An analysis of Eq.~(6) ($\xi=r/\Phi(t)$), Eq.~(7a)
($B_{\varphi}=\Phi^{\delta}Q_{\varphi}(\xi)$) and Eq.~(36a) leads to
the following
 expression:
$$
\frac{1}{r}\partial_r(rB_{\varphi})=0.\eqno(36b)
$$

Here, it is taken into account that
$\partial_r=\partial_\xi/\Phi(t)$.

 From Eq.~(36b) we can conclude that:
$$
B_{\varphi}=\frac{C}{r},\eqno(36c)
$$
with $C=\;$const.



Note that the expression for $B_{\varphi}$ is characterized by a
singularity at the axis ($r=0$). It seems reasonable to conclude
that, if we do not consider the axial stretching of self-similarly
evolving MCs, we can not obtain a physically valid solution for the
$B_{\varphi}-$ \textbf{component on the axis of the MC.}

\section{Numerical study: higher density and higher plasma $\beta$ case}

In this section, we investigate the evolution of MCs in a medium by
means of the model described in Section~2. For this purpose, the
Lagrangian numerical  MHD code \textbf{``Graale``} (Finn et al.\
\cite{fin04}) is used, which enables us to check whether the
above-obtained solutions maintain their self-similar nature when
they propagate in a medium. In the numerical code, azimuthal and
cylindrical symmetries are implied. Furthermore, it is assumed that
the magnetic structure expands uniformly in the longitudinal
direction, in other words $V_{z}=z\dot L/L;$ with $L$ the length of
the cylinder. An analysis of Eqs.~(20a-20c) and Eq.~(19) shows that
the derived expression for the longitudinal velocity coincides with
the one implemented in the code.

The assumptions of cylindrical and azimuthal symmetries in the code
and the prescription of the character of the longitudinal motions
makes the numerical simulations 1D. In the numerical runs the units
of the physical parameters are chosen as follows: the unit length
$L_{unit}=0.1AU=15\cdot10^{6}\;\textrm{km}$ is of the order of the
MC's radius at 1~AU, the unit magnetic field
$B_{unit}=3\;\textrm{nT}$, and the unit number density
$n_{unit}=10\;\textrm{cm}^{-3}$. After taking into account that the
proton mass $m_{p}\approx 1.7\cdot 10^{-27}\;\textrm{kg}$, one
derives that the unit mass density is
$\rho_{unit}=m_{p}n_{0}=0.8\cdot 10^{-14}\;\textrm{kg/m}^{3}$. The
unit speed is the Alfv\'en speed corresponding to $\rho_{unit}$ and
$B_{unit}$: $V_{unit}=V_{0A}= 20.5\;\textrm{km/s}$, which is of the
order of the MC's edges expansion velocity in the frame of the MC
(Vandas et al., \cite{van09}), and the unit time
$t_{unit}=\frac{L_{unit}}{V_{unit}}=200\;\textrm{h}$. A domain with
$R_{min}=0$ and $R_{max}=10$ is discretised with 2000 grid cells.
The time step used in the simulations is $\Delta t=5\cdot 10^{-7}$
($R_{min}$, $R_{max}$ and $\Delta t$ are given in units introduced
above-$L_{unit}$ and $t_{unit}$). Open boundary conditions are
applied. Inside the calculation domain, we introduce initial
conditions for the physical variables in two different regions:
inside the MC and outside the MC.

 Evidently, the solutions inside and outside the magnetic structure
should satisfy the following jump conditions across the surface of
any MC:
$$
[\varrho\upsilon_{r}]=0\eqno(37a)
$$
$$
\left[\varrho\upsilon_{r}^{2}+p+\frac{1}{8\pi}B^{2}\right]=0\eqno(37b)
$$
$$
\left[\frac{1}{2}\varrho\upsilon_{r}^{3}+(\frac{\gamma
p}{\gamma-1}+\frac{1}{4\pi}B^{2})\upsilon_{r}\right]=0\eqno(37c)
$$
and, finally,
$$
[\varrho\upsilon_{r}\upsilon_{t}]=0\eqno(37d)
$$

Here,  $[\cdot]$ denotes the jump of the quantity between the
brackets across the surface of the MC. Also,
$\upsilon_{r}=V_{r}-V_{s}$, where $V_{r}$ and $V_{s}$ are the plasma
and the MC's  surface velocity, respectively, while $\upsilon_{t}$
denotes the plasma velocity tangential to the surface of the MC.
Equation~(28a) and Eq.~(26) show that $\upsilon_{r}=0$, which is
logical for ideal MHD. Equations (37a,37c,37d) are satisfied for
arbitrary values of the plasma density and Eq.~(37b) leads to the
condition:
$$
\left[p+\frac{1}{8\pi}B^{2}\right]=0. \eqno(38)
$$

We know that the plasma mass density inside the MCs is lower than
outside them and the plasma $\beta$ within a MC is lower than in the
ambient plasma.  We therefore consider $\beta\sim 1$ in the ambient
environment and $\beta\sim 0.1$ inside the MC. For the magnetic
field within the MC, we use the solution expressed by
Eqs.~(33a-33c). For the magnetic field outside the MC, we assume
that the azimuthal component of the magnetic field $B_{\varphi
out}=0$, while the longitudinal component $B_{z out}$ is uniform. We
also assume that the mass density and the thermal pressure are
uniform in both regions of the computational domain.

Bearing in mind these assumptions and jump conditions $(37-38)$, one
can find explicit expressions for the plasma pressure and magnetic
field outside the MC.

Figure~1 represents the numerical solutions for the plasma mass
density and velocity, while Fig.~2 shows solutions for the magnetic
field components at different moments in time. On panels a and b we
plotted the dimensionless values of the density, velocity and
magnetic field.
 On panels c and d the dependence of the modified
values of the physical parameters on the self-similar variable is
presented.

The dependence of the modified mass density and velocity as well as the
dependence of the modified magnetic field components on the
self-similar variable clearly shows that our solutions maintain their
self-similarity in the course of the MC expansion.

\section{Discussion and Conclusions}

In this paper, we presented a detailed derivation of a class of
self-similar analytic solutions of the MHD equations for both
radially and axially expanding MCs and a numerical investigation of
these solutions. The usage of the self-similar approach is quite
common for the modeling of various kinds of solar plasma structures,
flows and eruptions (Low \cite{low82}, Osherovich \cite{osh93},
Farrugia et al. \cite{far95}, Nakwacki et al. \cite{nak08},
Shapakidze et al. \cite{shapo10}). In most of the previous studies,
however, \textit{only} the radial expansion of the MCs was
considered. In the present study, we took into  account also the
axial stretching of the MCs, which is a common observed feature of
at least some MCs. We have obtained explicit analytical expressions
for the magnetic field, the plasma velocity, the density and the
plasma pressure. Essentially, our solutions maintain their
self-similar nature during the whole course of their evolution and
propagation through the solar wind. These solutions are complete and
well-defined, fully analytic and, moreover, in the particular case
of the absence of the longitudinal expansion, our solutions
self-consistently match the analytic solutions derived by other
authors (Farrugia et al. \cite{far95}).

Note that for the class of solutions introduced by Eqs.~(7a-7d), the
assumptions of self-similarity and axial and azimuthal symmetry lead
to the fact that $\Phi(t)$ is a linear function of time ($\Phi(t)$
is the time dependent function of the self similar variable
$\xi=r/\Phi(t)$). In this case, the forces within the MCs are bound
to be balanced. We can thus conclude that the case in which the
magnetic structures are characterized by a low plasma-$\beta$,
corresponds to the force-free magnetic field case. Remark that this
result is also in agreement with the conclusion of \textbf{previous}
studies. We therefore believe that this is a correct and proper
time-dependent generalization of the widely used stationary
Lundquist model (Lundquist \cite{lun50}). Note also that in their
recent papers Vandas et al. (\cite{van06}, \cite{van09}) made a
comparison of the generalized Lundquist model with observations and
found good agreements between this classic model and the
experimental data.

It must be emphasized that our study is not the only one in which
the axial stretching of the MCs is taken into account together with
their radial expansion. As a matter of fact, Shimazu \& Vandas
(\cite{shi02}) also considered MCs with similar properties. In this
particular paper, the authors used the mathematical approach
introduced by Osherovich et al. (\cite{osh95}). In order to separate
the time-dependent parts of the solutions from multiplicative
functions of the self-similar variable only, Shimazu \& Vandas
(\cite{shi02}) imply a so-called ``separable magnetic'' field, which
was introduced in Osherovich et al.\ (\cite{osh95}). The approach
introduced in Osherovich et al.\ (\cite{osh95}), turns out to be
quite restrictive because it requires an \textit{ad hoc} relation
between the different components of the magnetic field. Also, in
order to separate the time-dependent part from the coordinate
dependent parts in the momentum equation, in addition to the
polytropic law, the authors introduced a specific mathematical
expression for the thermal pressure (see Eq.~17 Osherovich et al.\
\cite{osh95}). Actually, the mentioned expression relates pressure
and mass density (see Eqs.~(13-17) Osherovich et al.\ \cite{osh95}).
In our study, on the contrary, we used B.C.\ Low's approach
(\cite{low82}) and required a similar time-scaling for all parts of
the Lorentz force and the force caused by the gradient of the
thermal pressure. We argue that our approach is more general and
puts less non-physical restrictions upon the physical parameters.

Another difference of the results presented here with those of
Shimazu \& Vandas (\cite{shi02}) is with the temporal expansion
scaling. Shimazu \& Vandas assume that the longitudinal and the
radial expansion have the same time scaling, while in our work this
is \textit{not} assumed but it rather logically \textit{follows} as
the by-product of the accurate solution of the MHD equations. In
their paper, the time-dependent function of the self-similar
variable is characterized by a linear dependence on time
\textit{only} when the thermal pressure is zero, while we have
derived an explicit expression of this time-dependent function of
the self-similar variable and it has been shown that $\Phi$ is a
linear function of the time variable and it does not depend on the
character of the pressure function.

Yet another difference between the results presented here and the
results of Shimazu \& Vandas is related to the structure of the MC
magnetic field. Shimazu \& Vandas, in order to derive explicit
expressions for the magnetic field, used the assumption that the
magnetic structure of the MC is described by a force-free magnetic
flux rope. In our study, however, we derived an explicit expression
for the magnetic field. We have derived ordinary differential
equations (Eqs.~(30a)-(30b)) for the functions describing the
components of the magnetic field, after solving the equation of
motion for the case which corresponds to a low plasma $\beta$ within
the MC. For a particular type of parameters, we have found explicit,
analytical solutions for the components of the magnetic field
(Eqs.~(33a)-(33c)). Note that these expressions are a time dependent
generalization of the well-known Lundquist solutions (Lundquist
\cite{lun50}, Burlaga \cite{bur88}).



Our model implies the conservation of magnetic flux and helicity by
design, which is satisfactory and in good agreement with previous
investigations (Nakwacki et al. \cite{nak08}, Demoulin \& Dasso
\cite{dem09}, Kumar \& Rust \cite{kum96}).

For further confirmation of the validity of our solutions, we
investigated the dynamics of magnetic clouds numerically. In the
numerical code Graale we introduced our self-similar solutions as
initial conditions. The obtained numerical results showed that
during the evolution and propagation of these MCs, their physical
variables maintained their self-similar character. This circumstance
was illustrated by Fig.~1 and Fig.~2.

Obviously, the class of solutions found in this paper is quite
idealized. The assumptions about the self-similar evolution and the
consideration of a cylindrical symmetric structure are quite
well-justified, but real MCs show self-similar coherence and
cylindrical symmetry only approximately. Hence, in a future study
it would be reasonable and interesting to consider more realistic
configurations. There are several issues related to the model which
can be tested and generalized in a forthcoming study:
\begin{enumerate}

 \item
Our assumptions, just like in previous investigations (Low
\cite{low82}, Farrugia et al. \cite{far95}, Finn et al.
\cite{fin04}, Shapakidze et al. \cite{shapo10}), for the systems
where entropy is conserved, put a restriction on the value of the
polytropic index $\gamma=4/3$. We would like to develop a model that
helps to avoid this restriction.
 \item
Our model describes the plasma dynamics only inside the MC. In the near
future, we plan to investigate the interaction of an MC with its
environment by constructing consistent solutions of the MHD equations outside
the MC.
 \item
We investigated the obtained analytical solutions numerically using
a 1D MHD code and a simple model for the flow outside the MC was
implemented. It would be interesting and reasonable to study the MC
evolution also with 3D numerical codes, where a more complicated and
realistic background flow can be implemented (in preparation). For
this purpose the obtained solutions could be used as initial state
in the 3D numerical simulation codes. We are interested in an
investigation of the different possible boundary conditions on the
surface of the magnetic cloud.
\end{enumerate}

\begin{acknowledgements}

These results were obtained in the framework of the projects
GOA/2009-009 (K.U.Leuven), G.0304.07 (FWO-Vlaanderen) and C~90347
(ESA Prodex 9). Financial support by the European Commission through
the SOLAIRE Network (MTRN-CT-2006-035484), Georgian National Science
Foundation grant GNSF/ST06/4-096 and  funding from the European
Commission's Seventh Framework Programme (FP7/2007-2013) under the
grant agreement SOTERIA (project n� 218816, www.soteria-space.eu)
are gratefully acknowledged. AR acknowledges support of the Abdus
Salam ICTP through the Senior Associate Member reward and support of the ``Belgian Science
Policy'' (BELSPO) through a 2009 Fellowship To Non-EU Researchers.

\end{acknowledgements}

\appendix
\section{}
In this appendix we would like to give some details of derivation
\textbf{for}  Eq.~(21) and Eq.~(22).

During the process of derivation were implied following relations:

$$
\partial_r=\frac{1}{\Phi}\partial_\xi,\eqno(A1)
$$

$$
\partial_t=-\frac{\xi}{\Phi}\dot\Phi\partial_\xi.\eqno(A2)
$$

These relations follow from Eq.~(6).

Equation (21) is the radial component of the equation of
motion-Eq.~(4). In cylindrical coordinates the left term of Eq.~(4)
can be written as follows:
$$
\rho\left[\partial_t+(\textbf{V}\cdot\nabla)V_{r}\right]=\rho\left[\partial_t
V_{r}+V_{r}\partial_r V_{r}\right].\eqno(A3)
$$

If we take into account relation (6) and expression for $V_{r}$
(Eq.~11) we can obtain following expressions:

$$
\partial_t V_{r}=r
\frac{\ddot\Phi}{\Phi}-r\left(\frac{\dot\Phi}{\Phi}\right)^{2},\eqno(A4)
$$
$$
V_{r}\partial_r
V_{r}=r\left(\frac{\dot\Phi}{\Phi}\right)^{2}.\eqno(A5)
$$

After substitution of Eq.~(A4) and Eq.~(A5) into Eq.~(A3) we get:

$$
\rho\left[\partial_t+(\textbf{V}\cdot\nabla)V_{r}\right]=\rho\xi\ddot\Phi.\eqno(A6)
$$

If we combine of Eq.(7c) and Eq~(A6) we obtain:
$$
\rho\left[\partial_t+(\textbf{V}\cdot\nabla)V_{r}\right]=\Phi^{\alpha}\xi\ddot{\Phi}\tilde\rho,\eqno(A7)
$$
the left term of Eq.~(21).

In order to derive the first part of the right term of Eq.~(22) let
us introduce the following notation:

$$
\nabla\times\textbf{B}\equiv\textbf{J}.\eqno(A8)
$$

After taking into account expressions for magnetic field (7a-7b)
with $\sigma=-2$ we get:

$$
J_{r}=\frac{1}{r}\partial_\varphi B_{z}-\partial_z
B_{\varphi}=0,\eqno(A9)
$$

$$
J_{\varphi}=\partial_{z} B_{r}-\partial_r B_{z}=-\partial_r
B_{z}=-\frac{Q_{z}'}{\Phi^{3}}, \eqno(A10)
$$
and
$$
J_{z}=\frac{1}{r}\partial_r (r
B_{\varphi})-\frac{1}{r}\partial_\varphi B_{r}=\frac{1}{r}\partial_r
(rB_{\varphi})=\frac{Q_{\varphi}\Phi^{\delta-1}}{\xi}+Q_{\varphi}'\Phi^{\delta-1}.\eqno(A11)
$$

Since we know expressions of vector \textbf{J} we can derive the
vectorial product of \textbf{J} and \textbf{B} which represents the
first part of the right term of Eq.~(22).

 Combination of relation (A1)
with the expression for pressure Eq.~(7d) leads to the expression of
the second part of the right term of Eq.~(22).

\newpage

\begin{figure*}[]
\vspace*{1mm}
\begin{center}
\includegraphics[width=\textwidth]{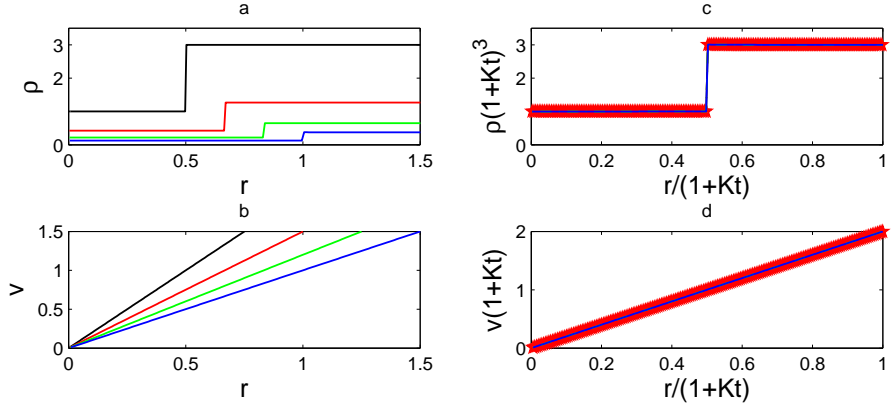}
\end{center}
\caption{Snapshots of the evolution of the plasma density and
velocity field. Panel~a) represents the dependence of the plasma
density on the radial coordinate for four moments in time; Panel~b)
shows the dependence on the plasma velocity on the radial coordinate
for four time moments; Panel~c) and Panel~d) illustrate the
dependence of the modified density and the modified velocity,
respectively, on the self similar variable. Parameter values for
this case are $k=2$, $B_{0}=1$, $\varrho_{out}=3\varrho_{in}$,
$\beta_{in}=0.1$, $\beta_{out}=1$, and $\Phi_{0}=1$. Black line
corresponds to $t=0$, red line represents the moment $t=0.5/3$,
green line shows the time moment $1/3$, and blue line corresponds to
the moment $t=0.5$.}
\end{figure*}

\newpage

\begin{figure*}[]
\vspace*{1mm}
\begin{center}
\includegraphics[width=\textwidth]{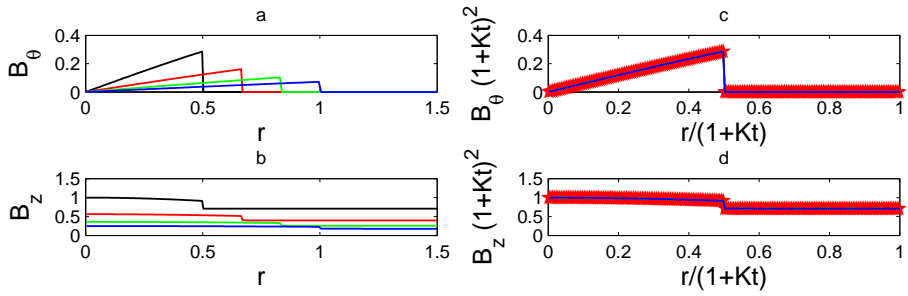}
\end{center}
\caption{Snapshots of the evolution of the magnetic field
components. Panel~a) represents the dependence of the azimuthal
component of the magnetic field on the radial coordinate for four
time moments; Panel~b) shows the dependence of the $z$-component of
the magnetic field on the radial coordinate for four time moments;
Panel~c) and Panel~d) illustrate the dependence of the modified
azimuthal and $z$-components of the magnetic field, respectively, on
the self similar variable. Parameter values for this case are $k=2$,
$B_{0}=1$, $\varrho_{out}=3\varrho_{in}$, $\beta_{in}=0.1$,
$\beta_{out}=1$, $r_{0}=1$, and $\Phi_{0}=1$. Black line corresponds
to $t=0$, red line represents the moment $t=0.5/3$, the line shows
the time moment $1/3$, and blue line corresponds to the moment
$t=0.5$. }
\end{figure*}

\end{document}